\documentclass[letterpaper,9pt,twocolumn,DIV=18]{scrartcl}
\usepackage[T1]{fontenc}
\usepackage[utf8]{inputenc}
\usepackage[font={small}]{caption}
\usepackage[font={small}]{subcaption}
\usepackage[bookmarks=false,hidelinks]{hyperref}
\usepackage{times}
\usepackage{amsmath,amssymb,amsthm}
\usepackage[affil-it]{authblk}
\usepackage{listings}
\usepackage[round]{natbib}
\usepackage{microtype}
\usepackage[shortlabels]{enumitem}
\usepackage{booktabs}
\usepackage{array}

\setkomafont{section}{\Large}
\setkomafont{subsection}{\large}
\addtokomafont{disposition}{\rmfamily}

\usepackage{graphicx}

\date{}

\usepackage[font={small}]{subcaption}
\usepackage{xspace}

\usepackage{xstring}

\usepackage{amsmath,amssymb}

\usepackage[ruled,vlined,linesnumbered]{algorithm2e}

\SetCommentSty{mycommentfont}
\DontPrintSemicolon
\SetKwComment{mycomment}{}{}

\usepackage{pgfplots}

\pgfplotsset{compat=newest}
\usetikzlibrary{plotmarks}
\usetikzlibrary{arrows.meta}
\usepgfplotslibrary{patchplots}
\pgfplotsset{every axis label/.append style={font=\tiny}}
\pgfplotsset{every tick label/.append style={font=\tiny}}

\usepackage{xspace}

\pgfkeys{/pgf/number format/.cd,1000 sep={\,}}

\newcommand{\addhistograms}[3]{\begin{axis}[at={(#2)},ybar,width=0.4\linewidth,height=0.3\linewidth,axis lines=left,ymin=0,ymax=1000,xmin=-0.1,xmax=.6,yticklabels={},xlabel=#3,x label style={at={(axis description cs:.98,-0.05)},anchor=west},]\addplot +[hist={bins=12,data min=0,data max=.5}, red!20!black,fill=red!80!white] table [y expr={min(1-\thisrowno{0},\thisrowno{0})}] {data/#1};\end{axis}}

\newcommand{\addhistogramslb}[5]{\begin{axis}[at={(#2)},ybar,width=#3,height=#4,axis lines=left,ymin=0,ymax=100,xmin=-0.1,xmax=.6,yticklabels={},xlabel={},x label style={at={(axis description cs:.98,-0.05)},anchor=west},]\addplot +[hist={bins=12,data min=0,data max=.5}, blue!20!black,fill=blue!80!white] table [y expr={min(1-\thisrowno{0},\thisrowno{0})}] {data/#1};\end{axis}}
\newcommand{\addhistogramsa}[5]{\begin{axis}[at={(#2)},ybar,width=#3,height=#4,axis lines=left,ymin=0,ymax=100,xmin=-0.1,xmax=0.6,yticklabels={},xlabel=#5,x label style={at={(axis description cs:.98,-0.05)},anchor=west},]\addplot +[hist={bins=12,data min=0,data max=.5}, red!20!black,fill=red!80!white] table [y expr={min(1-\thisrowno{0},\thisrowno{0})}] {data/#1};\end{axis}}

\DeclareMathOperator{\Var}{Var}
\DeclareMathOperator*{\argmin}{arg\,min}
\newcommand{\what}[1]{\hat{#1}}
\newcommand{\giv}[1]{\underline{#1}}

\renewcommand{\mid}{\,\vert\,}
\newcommand{\Prbs}{\mathbb{P}}
\newcommand{\Prb}[2]{\Prbs_{#1}\left({#2}\right)}
\newcommand{\Expb}[2]{\mathbb{E}_{#2}\left[{#1}\right]}
\newcommand{\Exp}[1]{\mathbb{E}\left[{#1}\right]}
\newcommand{\Ind}[1]{\mathbb{I}\left[{#1}\right]}

\newcommand{\y}{\mathbf{y}}
\newcommand{\z}{z}
\newcommand{\ysim}{\tilde{\y}}
\newcommand{\zsim}{\tilde{\z}}

\newcommand{\ygiv}{\giv{\smash{\y}}}
\newcommand{\zgiv}{\giv{\z}}

\newcommand{\weights}{w}
\newcommand{\param}{\boldsymbol{\theta}}
\newcommand{\paramset}{\Theta}

\newcommand{\parambest}{\boldsymbol{\theta}_{\hspace*{-.1em}\star}\hspace*{-.1em}}
\newcommand{\N}[1]{\mathcal{N}\left(#1\right)}
\newcommand{\U}{\mathcal{U}}
\newcommand{\lat}{\boldsymbol{\eta}}
\newcommand{\event}{S}

\newcommand{\pfa}{\text{\textsc{pfa}}}
\newcommand{\pfau}{\pfa_u}

\newcommand{\pfaus}{\pfa^\star_u}

\newcommand{\pfas}{\pfa^\star}
\newcommand{\apfas}{\hat{\pfa}^\star}
\newcommand{\apfau}{\hat{\pfa}_u}

\newcommand{\apfaus}{\hat{\pfa}_u^\star}

\newcommand{\dcc}{\textsc{Dcc}\xspace}

\newcommand{\MC}{\mathcal{P}_{\paramset}}

\title{Data Consistency Approach to Model Validation}

\author{Andreas Lindholm\thanks{\url{andreas.lindholm@it.uu.se}}} 
\author{Dave Zachariah\thanks{\url{dave.zachariah@it.uu.se}}} 
\author{Petre Stoica} 
\author{Thomas B. Sch\"on} 
\affil{Department of Information Technology, Uppsala University}
\publishers{\flushleft \textcolor{red}{Please cite the version published in}:\\\textit{IEEE Access}, 7(1):59788--59796, 2019.  DOI:10.1109/ACCESS.2019.2915109}

\begin{document}

\maketitle

\begin{abstract}
\itshape In scientific inference problems, the underlying statistical modeling assumptions have a crucial impact on the end results. There exist, however, only a few automatic means for validating these fundamental modelling assumptions. 
The contribution in this paper is a general criterion to evaluate the consistency of a set of statistical models with respect to observed data. This is achieved by automatically gauging the models' ability to generate data that is similar to the observed data.  Importantly, the criterion follows  from the model class itself and is therefore directly applicable to a broad range of inference problems with varying data types, ranging from independent univariate data to high-dimensional time-series. The proposed data consistency criterion is illustrated, evaluated and compared to several well-established methods using three synthetic and two real data sets.
\end{abstract}

\section{Introduction}
\label{sec:introduction}
In many scientific applications, statistical models provide a basis for inferences about real world phenomena. These inferences are typically dependent on the model assumptions being correct. However, there are few automatic means of evaluating these assumptions. In this paper, we address the problem of model validation by developing a method that automatically assesses the consistency of a set of models with respect to the observed data.

Let the observed data set be denoted as 
\begin{equation}
\ygiv = \{ \ygiv_1, \ygiv_2, \dots, \ygiv_n \},
\label{eq:dataset}
\end{equation}
which consists of $n$ data blocks of equal dimension, referred to as data points. We describe the mechanism that gave rise to the data by a probability density or mass function $p_0(\y)$, which is unknown to us. In many applications, the objective of statistical inference is to determine certain properties of the unknown function $p_0(\y)$. Statistical methods typically specify a family or class of probability distributions that aim to model $p_0(\y)$. We denote this \emph{model class} as 
$$\MC \triangleq \big\{ \: p(\y\mid\param):\param\in\paramset \: \big \},$$ 
where each model $p(\y\mid\param)$ is indexed by the parameter vector $\param$. Our aim in this paper is to assess whether the models in $\MC$ that best approximate $p_0(\y)$ are consistent with the observed data $\ygiv$ or not. The idea behind the proposed data consistency criterion (see below) is that if the best models in $\MC$ fail to generate data sets $\ysim$ that are `similar' to $\ygiv$, then $\MC$ can hardly be a valid modelling choice for $\ygiv$. This notion will be made precise in the subsequent sections.

Past research efforts have mostly focused on \emph{comparing} model classes, let us say $\MC'$ and $\MC''$, using tools such as the Akaike or Bayesian information criteria and Bayes factors \cite{Akaike:1974,Schwarz:1978,SM:2005,StoicaS:2004}. These criteria typically assume that some model class is well specified, meaning that one of the classes contains the unknown $p_0(\y)$. While the consistency criterion proposed in this paper also can be used for model comparison, our focus here is rather on \emph{validation} of \emph{one} specified model class $\MC$, which may or may not contain $p_0(\y)$.

An established approach to validation is to use a residual-based criterion which assesses whether there is any ``information'' left in the data after fitting a model. In the restricted context based on  the assumption of linear dynamical systems, such validation criteria are capable of rejecting model classes that are inconsistent with the data, cf. \cite{LB:78} and \cite[Ch. 11]{SoderstromStoica1988_system}. For nonlinear dynamical systems, there also exists a non-statistical validation method to assess whether a deterministic model is capable of matching the observed data \cite{Prajna:2005}. For statistical models based on the assumption of independent and identical distribution of the data points, classical tests such as Cram\'er-von Mises, Anderson-Darling and Kolmogorov-Smirnov tests are applicable \cite{AndersonD52,Lehmann:1975}. Those tests are, however, constructed for very specific model classes. In the context of Bayesian modelling, validation of a model class can also be performed using posterior predictive checks which require the user to specify a discrepancy measure, cf. \cite{R:84,GCS+:14,Box1980_sampling}.

Our proposed data consistency criterion (\dcc) evaluates the ability of a model to generate data similar to the observed one. In contrast to posterior predictive checks, $\dcc$ is automatic and does not require the user to specify any quantities except the model class $\MC$ itself. Furthermore, it applies directly to a broad range of model classes with various data types, e.g., linear regression models, count models, hidden Markov models, autoregressive models, etc. In general, $p(\y\mid\param)$ can be factored as $p(\y\mid\param) = \prod_{i=1}^n p(\y_i\mid \y_1,\dots, \y_{i-1},\param)$. \dcc is applicable whenever it is possible to point-wise evaluate $p(\y_i\mid \y_1,\dots, \y_{i-1},\param)$ for all $i$, and simulate new data $\ysim$ from the model, for any given $\param\in\paramset$. 

In summary, the proposed \dcc has the following features:

	\noindent $\bullet$ \dcc is a method for validating a given model class,\footnote{Unlike model selection methods, that performs a relative comparison between classes, e.g. Akaike and Bayesian information criteria.}
	
	\noindent $\bullet$ \dcc returns an interpretable test value,
	
	\noindent $\bullet$ \dcc follows automatically from the specified model class,
	
	\noindent $\bullet$ \dcc is readily applicable to a wide range of model classes, from models of univariate data points to high-dimensional time-series models.

The paper proceeds as follows: As an introductory application of our approach, we consider a modeling problem in seismology. Thereafter, in II, we explain the principles behind \dcc for a single model $p(\y\mid\parambest)$ and, subsequently in III, for an entire model class $\MC$. \dcc is then applied to the seismological problem as an illustration in IV. We also discuss its implementation and compare it to classical methods, and illustrate how it can be applied to some interesting model classes, including regression, autoregressive and latent variables models in IV. The source code for all experiments is available online.\footnote{\url{https://github.com/saerdna-se/consistency-criterion}}

\subsection*{Motivating example: Earthquake counts}

A standard assumption in earthquake analysis is that earthquakes occur independently as described by a Poisson point process. That is, the number of earthquakes in any given region during any given time interval is Poisson distributed. However, it is also well-known that earthquakes tend to be clustered (both in time and space), where each cluster typically has several `foreshocks' and `aftershocks' and one larger `mainshock'. By modeling the earthquakes within a cluster as a branching process, the negative binomial distribution has been suggested for earthquake counts \cite{Kagan14}. We consider both model classes, $\MC=\text{Poisson distribution}$ and $\MC=\text{negative binomial distribution}$, which have one and two free parameters, respectively. We will use our proposed method to assess whether these model classes are consistent with the data $\ygiv$ in the United States Geological Survey earthquake catalog\footnote{\url{https://earthquake.usgs.gov/earthquakes/search/}} (partly shown in Fig.~\ref{tab:eqdata}; the full data is found in Fig.~\ref{fig:data}). We will return to this example after developing the \dcc.

\begin{figure}[h]
	\centering
	\begin{tabular}{@{}ccccccc@{}}
		Magnitude & 2012 & 2013 & 2014 & 2015 & 2016 & 2017 \\ \hline
		$\geq$8 &2&2&1&1&0&1 \\
		$\geq$7 &16&19&12&19&16&7 \\
		$\geq$6 &133&142&155&146&146&111\\
		$\geq$5 &1680&1596&1729&1558&1696&1560 
	\end{tabular}
	\caption{A snippet of the global earthquake count data, for different magnitudes and years. Each row (magnitude class) is a different data set~$\ygiv$. We would like to assess the consistency between each of these data sets and the two model classes, the Poisson and negative binomial distributions, respectively.}
	\label{tab:eqdata}
\end{figure}

\section*{Data consistency check for a single model}

We begin by considering a model class consisting of only a single model, i.e., $\MC = \{ p(\y\mid\parambest)\}$ where $\parambest$ is a specified parameter.  Let $\ysim \sim p(\y\mid\parambest)$ denote a sample generated from the model and $\Prb{\ysim|\parambest}{\cdot}$ the probability of an event under the same model.

Initially, consider the simpler case of models in which the data points $i=1, \dots, n$ in \eqref{eq:dataset} are assumed to be independent.  Let $\zsim_i \triangleq \ln p(\ysim_i\mid\parambest)$ denote the log-likelihood for the $i$th generated data point, and let its mean be denoted as $\Exp{\zsim_i}$. For the $i$th observed data point, let  $\zgiv_i \triangleq \ln p(\ygiv_i\mid\parambest)$. The observed and generated log-likelihoods, $\zgiv_i$ and $\zsim_i$ form the basis of our criterion. Intuitively, if the deviation of  $\zgiv_i$ from $\Exp{\zsim_i}$ is much larger or much smaller than the deviation of $\zsim_i$ from $\Exp{\zsim_i}$, we consider the observed data $\ygiv$ to be \emph{atypical} for the given model $p(\y\mid\parambest)$. More formally, we define the following statistic
\begin{equation}\label{eq:T}
\begin{split}
T(\ygiv;\parambest) = \frac{1}{n} \sum_{i=1}^n \frac{\left(\zgiv_i - \Exp{\zsim_i}\right)^2}{\Var\left[\zsim_i\right]},
\end{split}
\end{equation}
where $\Var\left[\zsim_i\right]$ is variance of $\zsim_i$. Similarly, we define the statistic $T(\ysim;\parambest)$ for generated data by replacing $\ygiv$ with $\ysim$. Let us now define the random event $\event(\ysim,\ygiv)$ of generating a larger statistic than the observed one:
\begin{equation}
\begin{split}
\event(\ysim,\ygiv) \; : \; & T(\ysim;\parambest) > T(\ygiv;\parambest). 
\end{split}
\end{equation}
When the probability of this event $\Prb{\ysim|\parambest}{\event(\ysim,\ygiv)}$ is close to 0, it is highly improbable that $\ygiv$ could have been generated by $p(\y\mid\parambest)$ and we deem the model to be inconsistent with the observed data. See Fig.~\ref{fig1:b} for an illustration. This type of inconsistency is due to under-dispersion of the generated log likelihoods, compared to the observed ones. The probability of the complementary event $\Prb{\ysim|\parambest}{\event^c(\ysim,\ygiv)} = 1-\Prb{\ysim|\parambest}{\event(\ysim,\ygiv)}$ indicates inconsistency as well, see Fig.~\ref{fig1:c}. Also in this case it is improbable that $\ygiv$ could have been generated by $p(\y\mid\parambest)$. By contrast, if both aforementioned probabilities are significantly different from 0, we do not reject the model as inconsistent, see Fig.~\ref{fig1:a}.

The criterion above is readily generalized to models in which the data points in \eqref{eq:dataset} are dependent, by extending the definition of $z_i$ to
\begin{equation}\label{eq:z}
\z_i \triangleq \ln p(\y_i\mid\y_1, \dots, \y_{i-1}, \parambest).
\end{equation}
As above, the same symbols $\Exp{\zsim_i}$ and $\Var\left[\zsim_i\right]$ are used to define the mean and variance of $\zsim_i$.

If the model were a match of the unknown data-generating distribution, i.e., $p(\y\mid\parambest) = p_0(\y)$, then the quantity $\Prb{\ysim|\parambest}{\event(\ysim,\ygiv)}$ would be uniformly distributed between 0 and~1 with respect to the observed data $\ygiv$, see the proof below. In this situation, the probability of \emph{falsely} rejecting the model due to the generated log likelihoods being under-dispersed would be
\begin{equation}
\pfau(\parambest) \triangleq \Prb{\ysim|\parambest}{\event(\ysim,\ygiv)}.
\label{eq:pfau}
\end{equation}
Thus, when $p(\y\mid \param) = p_0(\y)$, the probability of $\pfau(\parambest)$ to be less than $\rho$, is equal to $\rho$.
Symmetrically, the false alarm probability due to over-dispersion is $1-\pfau(\parambest)$. If neither $\pfau(\parambest)$ nor $1-\pfau(\parambest)$ are small, we cannot reject the model $p(\y\mid\parambest)$ on the ground that the observed data $\ygiv$ is atypical, as discussed above. 

This criterion, that neither $\pfau(\parambest)$ nor $1-\pfau(\parambest)$ is close to~0, follows automatically from the specified model class and does not require user choices specific to the application scenario. The false alarm probabilities $\pfa$ above can be approximated numerically using Monte Carlo methods, as we will detail later.

\begin{figure}
	\centering
	\begin{subfigure}{.8\linewidth}
		\centering
		\input{figures/plot1}
		\caption{The observed data points appear as atypical with respect to the model, since they fall into regions of low probability. We obtain $T(\ygiv; \parambest) = 14$, and the probability of generating a higher statistic is $\Prb{\ysim|\parambest}{\event(\ysim,\ygiv)}= 0.00$. Thus, the dispersion of $\zsim_i$ is significantly lower than that of $\zgiv_i$ and we reject the model as inconsistent with the observed data.}
		\label{fig1:b}
	\end{subfigure}
	\begin{subfigure}{.8\linewidth}
		\centering
		\vspace{1em}
		\input{figures/plot3}
		\caption{The observed data points appear as atypical with respect to the model, since they are concentrated asymmetrically. We obtain $T(\ygiv; \parambest) = 0.24$, and the probability of generating a lower statistic is $\Prb{\ysim|\parambest}{\event^c(\ysim,\ygiv)}= 0.04$. Thus, the dispersion of $\zsim_i$ is significantly higher than that of $\zgiv_i$ and we reject the model as inconsistent with the observed data.}
		\label{fig1:c}
	\end{subfigure}
	\begin{subfigure}{.8\linewidth}
		\centering
		\vspace{1em}
		\input{figures/plot2}
		\caption{The observed data points appear to be typical samples from the model. We obtain $T(\ygiv; \parambest) = 0.51$, and the probability of generating a lower statistic is $\Prb{\ysim|\parambest}{\event^c(\ysim,\ygiv)}= 0.39$. We do therefore not reject the model as inconsistent.}
		\label{fig1:a}
	\end{subfigure}
	\caption{Consider a data set $\ygiv = \{ \ygiv_i \}$ containing $n=7$ two-dimensional data points (red triangles, upper panels) and assume a Gaussian i.i.d. model $p(\y\mid\parambest)=\prod_i p(\y_i\mid\parambest)$ (green level curves). The log-likelihoods $\zgiv_i = \ln p(\ygiv_i \mid \parambest )$ for each data point are shown as red triangles in the lower panels. The generated log-likelihoods $\zsim_i$ follow the distribution illustrated in green in the same panels. When the deviation of $\zgiv_i$ from $\Exp{\zsim_i}$ is significantly different from that of $\zsim_i$, it is unlikely that the model could have generated the observed sample. The deviation is quantified by the statistic $T(\ygiv; \parambest)$. The figure illustrates three cases: two cases of inconsistency (a, b) and a balanced case (c).}
	\label{fig1}
\end{figure}

\subsection*{Proof of uniform distribution of $\pfau$}

\newcommand{\xigiv}{\giv{\smash{\xi}}}
Let $\xi \triangleq T(\ysim;\parambest)$, whose distribution is characterized by a cumulative density function denoted $F_\xi(x)$. Further, let $\xigiv \triangleq T(\ygiv;\parambest)$, which allows us to write
\begin{align}
\Prb{\ysim|\parambest}{\event(\ysim,\ygiv)} = 
\Prb{\ysim|\parambest}{\xi > \xigiv} = F_\xi(\xigiv).
\end{align}
Now, if $\ygiv\sim p(\y\mid\parambest)$, then also the distribution of $\xigiv$ is characterized by $F_\xi$, implying that $F_\xi(\xigiv)\sim\mathcal{U}[0,1]$ according to the probability integral transform \cite[Thm. 2.1.10]{Casella&Berger2002_statistical}.

\newpage
\section*{Data consistency check for the best models in a class}

In most applications, $\parambest$ is not given. Instead $\MC$ may consist of a large number of models, possibly an uncountable number when $\param\in\Theta$ is continuous. In such a case, we will aim to evaluate the false alarm probabilities $\pfau(\param)$ and $1-\pfau(\param)$ with respect to the models $p(\y\mid \param)$ that best approximate $p_0(\y)$. A natural measure for quantifying the accuracy of the approximation is the Kullback-Leibler divergence \cite{Kullback1997_information}. The best models are then defined as those that minimize the divergence:
\vspace{-.5em}
\begin{equation}
\parambest \: \in \: \argmin_{\param} \; \underbrace{\Expb{ \ln p_0(\y) - \ln p(\y | \param)  }{0}}_{\text{model divergence}},
\label{eq:mindivergencemodels}
\end{equation}
\vspace{-.5em}

\noindent 
where the expectation $\mathbb{E}_0$ is with respect to $\y\sim p_0(\y)$. In particular, if $\parambest$ exists such that the model divergence attains the minimum value $0$, it follows that $p(\y\mid\parambest)=p_0(\y)$.

Since $p_0(\y)$ is unknown, \eqref{eq:mindivergencemodels} cannot be used to identify the best models. Consequently we resort to an alternative approach. We assign weights to each model in $\MC$ so as to average the false alarm probability $\pfau(\param)$ across those models that are likely to be the best approximations of $p_0(\y)$. The averaged false alarm probability due to under-dispersion is given by
\begin{equation}
\boxed{
	\pfaus = \int_\Theta \pfau(\param) \weights(\param \mid \ygiv) \: d\param
}
\label{eq:dcce}
\end{equation}
where the weights $\weights(\param\mid\ygiv) \geq 0$ are high for models in the neighborhood of $\parambest$ and integrate to unity. By defining
\begin{equation}
\weights(\param \mid \ygiv ) \triangleq \frac{\weights_0(\param) p(\ygiv\mid\param)}{\int_\Theta \weights_0(\param) p(\ygiv\mid\param) d\param},
\label{eq:weights}
\end{equation}
the weights reflect the uncertainty about the location of $\parambest$ in the parameter space $\Theta$ \cite{Casella&Berger2002_statistical,BissiriEtAl2012_converting}. The default choice of the initial weights is $w_0(\param) \equiv $ 1. In certain applications, however, we may have prior information about the location of $\parambest$ in $\Theta$. Then the initial weights $\weights_0(\param)$ can be chosen to describe these prior beliefs. Under certain regularity conditions, the weights in \eqref{eq:weights} concentrate at $\parambest$ as $n \rightarrow \infty$ if $\parambest$ is unique, cf. \cite{LC:79} and \cite{B:66}.

We denote our final criterion
\begin{equation}
\boxed{
	\pfas = \min\left(\pfaus,1-\pfaus\right)}
\label{eq:dcc}
\end{equation}

In summary, the proposed data consistency criterion (\dcc) for the model class $\MC$ is the minimum of $\pfaus$ and $1-\pfaus$ \eqref{eq:dcce}. A value close to $0$ indicates that the observed data~$\ygiv$ is atypical for the best models in $\MC$, and thereby we consider the model class $\MC$ to be inconsistent with $\ygiv$.

\subsection*{Implementation}
The averaged $\pfas$ is available in closed-form only in very few special cases, since there are non-trivial integrals in \eqref{eq:pfau}, (\ref{eq:dcce}), and (\ref{eq:weights}). However, the integrals can be efficiently approximated by Monte Carlo integration techniques, provided that data $\ysim$ can be generated from $p(\y\mid\param)$ and that $p(\y_i\mid \y_{1}, \dots, \y_{i-1},\param)$ can be evaluated point-wise. We outline such a generic Monte Carlo-based implementation
in Algorithm~\ref{alg:impl}, where $N$ parameters are first drawn using $w(\param\mid\y)$, and then (for each such draw) $M+M'$ samples of $\ysim$ are generated from $p(\y\mid\param)$, giving a computational complexity on the order of $N(M+M')$. 

The operations needed to execute Algorithm~\ref{alg:impl} are common in most statistical software packages. The weights $w(\param\mid\ygiv)$ can be computed (at least approximately) by methods that estimate or learn $\param$. Numerical evaluation of the (incremental) likelihoods $p(\y_i\mid \y_{1}, \dots, \y_{i-1},\param)$, which in Algorithm~\ref{alg:impl} has to be performed both for generated data $\ysim$ (line 5) and observed data $\ygiv$ (line 6), is possible for many statistical models. If $n$ is large compared to the number of parameters, it can be justified to use the following weights
\vspace{-.8em}
$$w(\param\mid\ygiv) = \begin{cases} 1, & \param = \what{\param} \\ 0, & \param \neq \what{\param}  \end{cases},$$
\vspace{-1em}

\noindent where $\what{\param}$ is the maximum likelihood/maximum a posteriori point estimate (in this case $N=1$). 

\begin{algorithm}[t]\small
		Construct $\weights(\param\mid\ygiv)$\;
		Draw $N$ samples $\param^{(j)}\sim \weights(\param\mid\ygiv)$, $j=1, \dots, N$\;
		\For{$j = 1, \dots, N$}{
			Simulate $M'$ data sets $\ysim^{\prime (k)}\sim p(\y\mid\param^{(j)})$, $k=1, \dots, M'$\;
			Compute $\zsim_i^{\prime(k)}$ as \eqref{eq:z} for all generated data points $\ysim^{\prime (k)}_i$\;
			Compute sample mean $\hat m_i$ and variance $\hat v_i$ of $\zsim'_i$, $i=1, \dots, n$\;
			Simulate $M$ data sets $\ysim^{(\ell)}\sim p(\y\mid\param^{(j)})$, $\ell=1, \dots, M$\;
			Compute $\zsim_i^{(\ell)}$ as \eqref{eq:z} for all generated data points $\ysim^{(\ell)}_i$\;
			Compute $\zgiv_i$ as \eqref{eq:z} for all observed data points $\ygiv_i$\;
			Compute $\giv{T}^{(j)} = \frac{1}{n}\sum_{i=1}^n \frac{\left(\zgiv_i-\hat m_i\right)^2}{\hat v_i}$\;
			Compute $\tilde{T}^{(j,\ell)} = \frac{1}{n}\sum_{i=1}^n \frac{\left(\zsim_i^{(\ell)}-\hat m_i\right)^2}{\hat v_i}$\;
			Set $\apfau^{(i)} = \frac{1}{M}\sum_{\ell=1}^M\Ind{\tilde{T}^{(j,\ell)}>\giv{T}^{(j)}}$\;
		}
		Set $\apfaus = \frac{1}{N}\sum_{i=1}^N\apfau^{(i)}$ and $\apfas = \min\left(\apfaus,1-\apfaus\right)$\;
		\caption{Monte Carlo implementation of DCC \eqref{eq:dcce}}
		\label{alg:impl}
\end{algorithm}

\section{Examples}

\subsection*{Earthquake counts (cont'd)}
To assess whether the Poisson or the negative binomial distribution is best suited for the data (partly) presented in Fig.~\ref{tab:eqdata}, we are now ready to apply the \dcc. We use the Monte Carlo approach in Algorithm~\ref{alg:impl} with $N = 200$ and $M=M' = 200$, and obtain the results in Fig.~\ref{tab:eqres}. From this table we draw the conclusion that the Poisson distribution and the negative binomial distribution are both consistent with the data for earthquakes with magnitude $\geq 7$. However, only the negative binomial distribution is  consistent with the data for smaller magnitudes. This result supports the qualitative reasoning in the literature \cite{Kagan14}: the Poisson distribution does not take the clustering effects into account, but since each cluster typically does not contain more than one major earthquake, the number of really big earthquakes can still follow the Poisson distribution.

\begin{figure}[h]
	\centering\small
	\begin{tabular}{@{}ccc@{}}
		Magnitude & Poisson distribution & Negative binomial distribution \\ \hline
		$\geq$8 & $\apfas=0.40$ & $\apfas=0.39$ \\
		$\geq$7 & $\apfas=0.29$ & $\apfas=0.38$ \\
		$\geq$6 & $\apfas=0.00$ & $\apfas=0.30$ \\
		$\geq$5 & $\apfas=0.00$ & $\apfas=0.13$ \\ \hline
	\end{tabular}
	\caption{Assessment of the two earthquake count models (Poisson and negative binomial) for earthquakes of different magnitudes. The low average false alarm probabilities $\apfas$ for the Poisson distribution suggests that this model class is not consistent with the observed data for magnitudes $\leq 6$.}
	\label{tab:eqres}
	\vspace{-1em}
\end{figure}


\subsection*{Synthetic data: Gaussian models}

\begin{figure}[t]
	\centering
	\begin{subfigure}{.9\linewidth}
		\centering
		\begin{tikzpicture}
		\addhistograms{correctmodel10.csv}{.00\linewidth,.36\linewidth}{$\apfas$}
		\addhistograms{correctclass10.csv}  {.4\linewidth,.36\linewidth}{$\apfas$}
		\addhistograms{correctmodel100.csv}{.00\linewidth,.18\linewidth}{$\apfas$}
		\addhistograms{correctclass100.csv}  {.4\linewidth,.18\linewidth}{$\apfas$}
		\addhistograms{correctmodel1000.csv}{.00\linewidth,.0\linewidth}{$\apfas$}
		\addhistograms{correctclass1000.csv}  {.4\linewidth,.0\linewidth}{$\apfas$}

		\node (sigma01) at (.12\linewidth,-.75) {\footnotesize$\MC=\{\N{0,1}\}$};
		\node (sigma1) at (.56\linewidth,-.75) {\footnotesize$\MC'=\{\N{\mu,\sigma^2}\}$};
		
		\node[rotate=90] (n10) at (-.4,.42\linewidth) {\footnotesize$n=10$};
		\node[rotate=90] (n100) at (-.4,.24\linewidth) {\footnotesize$n=100$};
		\node[rotate=90] (n1000) at (-.4,.06\linewidth) {\footnotesize$n=1\,000$};
		\end{tikzpicture}
		\vspace{-.5em}
		\caption{Data generating process $p_0 = \N{0,1}$. Both model classes contain $p_0$. As $n$ increases, $\pfas$ approaches a uniform distribution also for $\MC'$.}
		\label{fig:simplea}
	\end{subfigure}

	\begin{subfigure}{.9\linewidth}
		\centering
		\begin{tikzpicture}
		\addhistograms{wrongmodel10.csv}{.00\linewidth,.36\linewidth}{$\apfas$}
		\addhistograms{wrongclass10.csv}  {.4\linewidth,.36\linewidth}{$\apfas$}
		\addhistograms{wrongmodel100.csv}{.00\linewidth,.18\linewidth}{$\apfas$}
		\addhistograms{wrongclass100.csv}  {.4\linewidth,.18\linewidth}{$\apfas$}
		\addhistograms{wrongmodel1000.csv}{.00\linewidth,.0\linewidth}{$\apfas$}
		\addhistograms{wrongclass1000.csv}  {.4\linewidth,.0\linewidth}{$\apfas$}

		\node (sigma01) at (.12\linewidth,-.75) {\footnotesize$\MC=\{\N{0,1}\}$};
		\node (sigma1) at (.56\linewidth,-.75) {\footnotesize$\MC'=\{\N{\mu,\sigma^2}\}$};
		
		\node[rotate=90] (n10) at (-.4,.42\linewidth) {\footnotesize$n=10$};
		\node[rotate=90] (n100) at (-.4,.24\linewidth) {\footnotesize$n=100$};
		\node[rotate=90] (n1000) at (-.4,.06\linewidth) {\footnotesize$n=1\,000$};
		\end{tikzpicture}
		\vspace{-.5em}
		\caption{Data generating process $p_0 = \U[0,1]$. Neither model class contain $p_0$. As $n$ increases, $\pfas$ concentrates at 0 and both model classes are rejected.}
		\label{fig:simpleb}
	\end{subfigure}
\vspace{-.5em}
	\caption{Histograms of \dcc obtained from 1000 experiments, when $p_0$ is the standard Gaussian distribution (a) and the standard uniform distribution (b). Two different model classes are considered in the left and right columns, respectively. \dcc is able to correctly identify the cases where the model class is consistent/inconsistent with the data.}
	\label{fig:simple}
	\vspace{-1em}
\end{figure}
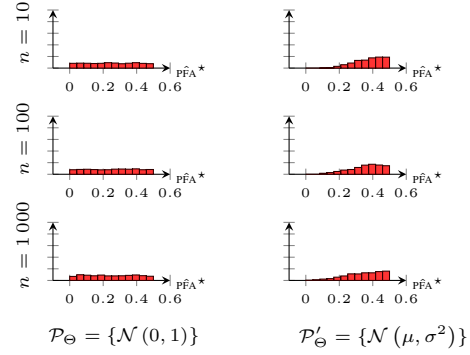
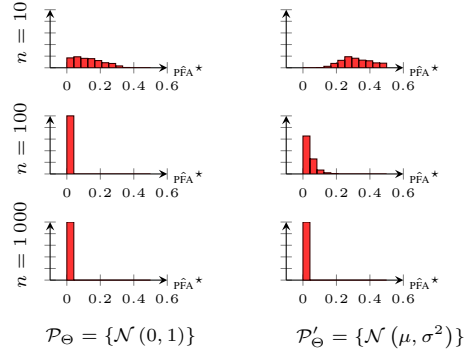

To further illustrate the behavior of  $\pfas$, we conduct a simulation study with two Gaussian model classes $\MC = \{\N{0,1}\}$ and $\MC' = \{\N{\mu,\sigma^2}: \sigma^2 > 0\}$ with $\param=\{\mu,\sigma^2\}$. Note that  $\MC$ contains only a single model. We consider data sets $\ygiv \sim p_0(\y)$ of different sizes $n$ from a standard Gaussian distribution $p_0 = \N{0,1}$ and a standard uniform distribution $p_0 = \U[0,1]$, respectively. We use $N=50$ and $M = M' = 100$, and evaluate \dcc in $1\,000$ experiments. The results are summarized as histograms in Fig.~\ref{fig:simple}.

In the case when data comes from $p_0 = \N{0,1}$ (Fig.~\ref{fig:simplea}), both $\MC$ and $\MC'$ contain $p_0$. Furthermore, the only model in $\MC$ is $p_0$. Thus for $\MC$ the averaged $\pfas$ is uniformly distributed across experiments (cf. the proof of uniform distribution for $\pfa$). In contrast to this, the weights $\weights(\param\mid \ygiv)$ for $\MC'$ concentrate at the best model $p_0$ only as $n\to\infty$. Thus the averaged $\pfas$ approaches a uniform distribution only asymptotically. Neither model class is falsely rejected more frequently than the $\pfas$ indicates.

In the case when data comes from $p_0 = \U[0,1]$ (Fig.~\ref{fig:simpleb}), neither $\MC$ nor $\MC'$ contains $p_0$. Unlike $\MC$, however, the best model in $\MC'$ matches the mean and variance of $p_0$. Therefore, the inconsistency of the best model is discernible only for data points generated from the distribution tails. Consequently, it takes more samples $n$ to reject the larger model class $\MC'$ than it takes to reject $\MC$. As $n$ increases, both model classes are clearly rejected.

\newpage
Using the same example, we next make a comparison to the classical Kolmogorov-Smirnov \cite{Massey51}, Lilliefors \cite{Lillienfors67}, Anderson-Darling \cite{AndersonD52}, and Jarque-Bera \cite{JarqueB87} tests of normality. We consider $n=100$, and select the respective thresholds such that the rate of falsely rejecting the model is either $5\%$ or $10\%$. The Kolmogorov-Smirnov test is by construction a test against $\MC = \N{0,1}$, whereas the others are tests against $\MC' = \N{\mu,\sigma^2}$. To achieve 5\% or 10\% false rejection probability in the classical tests, tabulated threshold values are available.  To select the \dcc threshold value for $\MC$ (no unknown parameters), the uniform distribution result can be directly applied, meaning that a false rejection rate of $\rho$ is achieved by selecting the threshold\footnote{The factor 1/2 is because $\pfau$ is uniformly distributed on $U[0,1]$, which implies that $\min(\pfau,1-\pfau)$ is uniformly distributed on half the interval, $[0,0.5]$.} as $\rho/2$. In the case of $\MC'$ with unknown parameters, however, the same procedure would be overly conservative (as indicated by Fig.~\ref{fig:simplea}), since the uniform property is only asymptotic as $n\to\infty$. Instead, the threshold is set using simulations: Data with $n=100$ is repeatedly simulated from $\N{0,1}$ (which lies in $\MC'$, hence a consistent case), and a threshold is selected such that the desired false rejection rate is achieved (cf. second column in Fig.~\ref{tab:gauss}). In fact, a threshold selected in this fashion will be independent of the parameters used in the simulations, since different parameter values essentially only change a constant in both $\zgiv_i$ and $\zsim_i$. This simulation-based procedure for selecting \dcc thresholds for model classes with unknown parameters is hence useful also for practical purposes.

\begin{figure}[t]
	\centering

	\begin{subfigure}{\linewidth}
		\resizebox{\linewidth}{!}{\begin{tabular}{@{}p{1.3cm}>{\centering\arraybackslash}p{2cm}>{\centering\arraybackslash}p{2.3cm}>{\centering\arraybackslash}p{2cm}>{\centering\arraybackslash}p{2.3cm}@{}}
				Method &\small $\MC=\N{0,1}$\newline$p_0=\N{0,1}$ &\small  $\MC'=\N{\mu,\sigma^2}$\newline$p_0=\N{0,1}$ & \small $\MC=\N{0,1}$\newline$p_0=\U[0,1]$ & \small $\MC'=\N{\mu,\sigma^2}$\newline$p_0=\U[0,1]$ \\ \midrule
				\textit{Oracle} & \textit{0} & \textit{0} & \textit{1} & \textit{1} \\ \midrule
				DCC & 0.10 & 0.10 & 1.00 & 1.00 \\ \midrule
				Kolmogorov-Smirnov & 0.10 & - & 1.00 & - \\ \midrule
				Anderson-Darling & - & 0.09 & - & 1.00 \\ \midrule
				Lilliefors & - & 0.09 & - & 0.77 \\ \midrule
				Jarque-Bera & - & 0.10 & - & 0.99 \\
		\end{tabular}}
		\caption{10\% false rejections}
	\end{subfigure}

	\vspace{1em}
	
	\begin{subfigure}{\linewidth}
		\resizebox{\linewidth}{!}{\begin{tabular}{@{}p{1.3cm}>{\centering\arraybackslash}p{2cm}>{\centering\arraybackslash}p{2.3cm}>{\centering\arraybackslash}p{2cm}>{\centering\arraybackslash}p{2.3cm}@{}}
				Method &\small $\MC=\N{0,1}$\newline$p_0=\N{0,1}$ &\small  $\MC'=\N{\mu,\sigma^2}$\newline$p_0=\N{0,1}$ & \small $\MC=\N{0,1}$\newline$p_0=\U[0,1]$ & \small $\MC'=\N{\mu,\sigma^2}$\newline$p_0=\U[0,1]$ \\ \midrule
				\textit{Oracle} & \textit{0} & \textit{0} & \textit{1} & \textit{1} \\ \midrule
				DCC & 0.06 & 0.05 & 1.00 & 1.00 \\ \midrule
				Kolmogorov-Smirnov & 0.05 & - & 1.00 & - \\ \midrule
				Anderson-Darling & - & 0.04 & - & 0.96 \\ \midrule
				Lilliefors & - & 0.04 & - & 0.60 \\ \midrule
				Jarque-Bera & - & 0.05 & - & 0.75 \\
		\end{tabular}}
		\caption{5\% false rejections}
	\end{subfigure}
	\caption{Probability of rejection, estimated using 1000 simulations. For each method, the rejection thresholds are set to achieve a certain probability of falsely rejecting the model class (i.e., the first and second columns). The proposed \dcc clearly performs on par with the classical methods, even though \dcc is a much more generally applicable method. The ``oracle'' is a fictitious test which knows the ground truth.}
	\label{tab:gauss}
\end{figure}

We simulate 1000 experiments, and report the estimated probability of rejections in Fig.~\ref{tab:gauss}. The results suggest that \dcc has a performance comparable to the performance of the considered classical tests. In fact, \dcc even outperforms the Lilliefors and Jarque-Bera tests for this particular example. It should, moreover, be remembered that whereas these classical tests are essentially tailor-made for testing normality of scalar random variables, the proposed \dcc is generally applicable also to much more complex models, such as multivariate non-linear time-series models.

\subsection*{Synthetic data: Regression models}
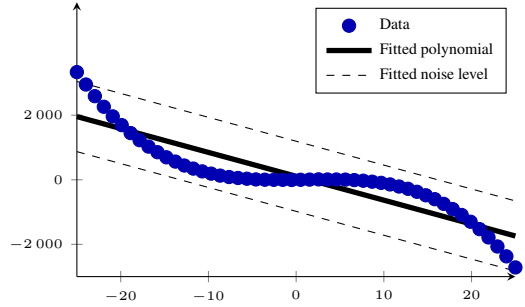
\begin{figure}[t]
	\centering
%
%

\begin{tikzpicture}

\begin{axis}[%
width=.85\linewidth,
height=.6\linewidth,
xmin=-25,
xmax=25,
xtick={-20, -10,   0,  10,  20},
ymin=-3000,
ymax=5500,
ytick={-2000,     0,  2000},
legend style={legend cell align=left, align=left, font=\scriptsize},
axis lines=left
]
\addplot [color=blue, only marks, mark size=2.5pt, mark=*, mark options={solid, blue!70!black}]
  table[row sep=crcr]{%
-25	3336.20\\
-23.97	2951.70\\
-22.95	2590.67\\
-21.93	2262.55\\
-20.91	1964.10\\
-19.89	1692.86\\
-18.87	1447.00\\
-17.85	1227.21\\
-16.83	1028.55\\
-15.81	854.29\\
-14.79	696.39\\
-13.77	564.20\\
-12.75	442.33\\
-11.73	346.84\\
-10.71	260.04\\
-9.69	190.73\\
-8.67	129.48\\
-7.65	85.77\\
-6.63	56.17\\
-5.61	32.71\\
-4.59	10.98\\
-3.57	2.40\\
-2.55	-0.06\\
-1.53	-0.68\\
-0.51	-5.61\\
0.51	0.04\\
1.53	5.74\\
2.55	12.94\\
3.57	8.81\\
4.59	8.96\\
5.61	5.38\\
6.63	-9.69\\
7.65	-28.84\\
8.67	-58.83\\
9.69	-94.85\\
10.71	-143.87\\
11.73	-208.37\\
12.75	-279.07\\
13.77	-371.65\\
14.79	-480.45\\
15.81	-601.76\\
16.83	-746.54\\
17.85	-908.29\\
18.87	-1090.53\\
19.89	-1300.16\\
20.91	-1528.91\\
21.93	-1785.80\\
22.95	-2066.44\\
23.97	-2372.47\\
25	-2715.46\\
};
\addlegendentry{Data}

\addplot [color=black, line width=2.0pt]
  table[row sep=crcr]{%
-25	1958.66\\
25	-1741.99\\
};
\addlegendentry{Fitted polynomial}

\addplot [color=black, dashed]
  table[row sep=crcr]{%
-25	3045.31\\
25	-655.34\\
};
\addlegendentry{Fitted noise level}

\addplot [color=black, dashed, forget plot]
  table[row sep=crcr]{%
-25	872.01\\
25	-2828.64\\
};
\end{axis}
\end{tikzpicture}%
	\caption{Data points from $p_0$ (3rd order polynomial, blue dots) and model class $\MC = \{ \text{1st order polynomial} + \text{Gaussian noise} \}$ where $\param$ contains the polynomial coefficients and noise variance. A fitted model is shown with $\param$ estimated using the maximum likelihood method (black solid line: estimated polynomial, dashed lines: 2 estimated noise standard deviations). While the estimated noise variance produces a good fit in terms of likelihoods, this model class should ideally be rejected. Using \dcc for $\MC$, we obtain $\pfas = 0.01$ and can reject $\MC$. Similarly, for a model class $\MC'$ containing 2nd order polynomials we obtain $\pfas=0.01$. By contrast, for 3rd order polynomials, $\MC''$, we have $\pfas = 0.37$ and this class is correctly found not to be inconsistent with the data.}
	\label{fig:poly}
\end{figure}

We illustrate the capability of \dcc to reject model classes with an inappropriate model order, using the example of polynomial regression. If the observed data is well described by a high-order polynomial, the best model in $\MC$, which contains models of lower-order polynomials plus noise, will yield a good fit in the likelihood sense because the noise variance is scaled to match the residuals. Such an example is shown in Fig.~\ref{fig:poly}, where $\MC$ contains 1st order polynomial models with independent Gaussian noise. When assessing $\MC$ using \dcc, we obtain $\apfas = 0.01$ (with  $N=M=M'=100$), which clearly indicates an inconsistency. On the other hand, for the model class containing 3rd order polynomials we obtain $\apfas = 0.37$, which (correctly) indicates no inconsistency.

\subsection*{Synthetic data: Time-series models}
For the case of time-series models driven by white noise, whiteness tests such as the Ljung-Box test \cite{LB:78,Stoica:1977} are common validation techniques. The Ljung-Box test constructs a p-value from the fact that the statistic $n(n+2)\sum_{k=1}^h\frac{\hat{r}_k^2}{n-k}$ will follow a $\chi^2_{h-d}$ distribution if the model is correct, with $\hat{r}_k$ being the lag $k$ sample correlation of the prediction residuals, and $d$ the dimension of $\param$. We set the upper lag limit $h$ to $\log n$ (rounded to nearest integer).

We conduct a simulation study with data from the saturated first order autoregressive model $y_i = \max( \: 0.7 y_{i-1} + e_i , \: -0.3 \: )$, where $e_i\sim\N{0,1}$. We assume a misspecified model class which consists of first-order linear autoregressive models $\MC = \{p(\y|\param) : y_i =a y_{i-1} + e_i, \: e_i\sim\N{0,\sigma^2}: \sigma^2 > 0\}$, with unknown parameters $\theta = \{a , \sigma^2 \}$. We consider cases where $\ygiv$ contains different amount of data samples $n$, and use $N=200$ and $M=M'=200$ in Algorithm~\ref{alg:impl}.

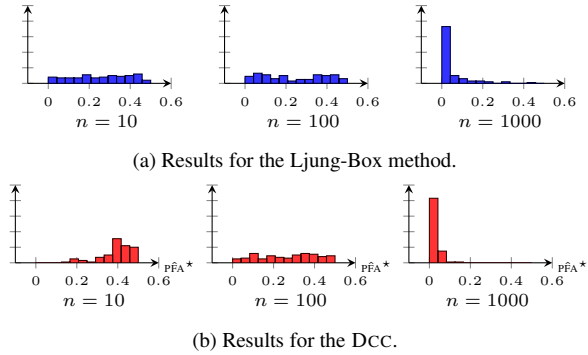
\begin{figure}[t]
	\begin{subfigure}{\linewidth}
		\centering
		\begin{tikzpicture}
		\addhistogramslb{lb_1_10.csv}{.00\linewidth,0}{.4\linewidth}{.3\linewidth}{}
		\addhistogramslb{lb_1_100.csv}{.3\linewidth,0}{.4\linewidth}{.3\linewidth}{}
		\addhistogramslb{lb_1_1000.csv}{.6\linewidth,0}{.4\linewidth}{.3\linewidth}{}
		
		\node (n1) at (.12\linewidth,-0.5) {\footnotesize$n=10$};
		\node (n2) at (.42\linewidth,-0.5) {\footnotesize$n=100$};
		\node (n3) at (.72\linewidth,-0.5) {\footnotesize$n=1000$};
		\end{tikzpicture}
		\caption{Results for the Ljung-Box method.}
	\end{subfigure}
	\begin{subfigure}{\linewidth}
		\centering
		\begin{tikzpicture}
		\addhistogramsa{ARwhitecomp10.csv}{.00\linewidth,0}{.4\linewidth}{.3\linewidth}{$\apfas$}
		\addhistogramsa{ARwhitecomp100.csv}{.3\linewidth,0}{.4\linewidth}{.3\linewidth}{$\apfas$}
		\addhistogramsa{ARwhitecomp1000.csv}{.6\linewidth,0}{.4\linewidth}{.3\linewidth}{$\apfas$}
		
		\node (n1) at (.12\linewidth,-0.5) {\footnotesize$n=10$};
		\node (n2) at (.42\linewidth,-0.5) {\footnotesize$n=100$};
		\node (n3) at (.72\linewidth,-0.5) {\footnotesize$n=1000$};
		\end{tikzpicture}
		\caption{Results for the \dcc.}
	\end{subfigure}
	\caption{Histograms for the Ljung-Box p-value (top) and $\apfas$ (bottom). For both methods, small values indicate an inconsistency. The data $\ygiv$ is generated by a saturated (non-linear) autoregressive model, but the assumed model class $\MC$ consists of linear autoregressive models. In this example, the amount of data required by the Ljung-Box method to detect the mismatch is slightly larger than for the proposed criterion.}
	\label{fig:simple2}
\end{figure}

The results are shown in Fig.~\ref{fig:simple2}. As the amount of data $n$ grows, both methods correctly reject the misspecified model, with the proposed \dcc needing slightly less data to do so than the much more specialized Ljung-Box method.

\subsection*{Latent-variable models: Evolution of a kangaroo population}

Certain models are too complex to be described using closed-form expressions. Instead these models are often parameterized using latent variables~$\lat$, with a prior distribution $p(\lat\mid\param)$. This includes, e.g., hidden Markov or state-space models, mixed-effect models, latent topic models, etc. Then the data distribution can be written as the integral 
\begin{align}
p(\y \mid \param )= \int p(\y\mid\lat,\param)p(\lat\mid\param)d\lat.
\end{align}
If the (incremental) likelihoods $p(\y_i\mid\y_{1}, \dots, \y_{i-1},\param)$ can be evaluated or well approximated, \dcc can be computed also for this model class, as we will illustrate in the following.

We consider the dynamics of a population of red kangaroos (\textit{Macropus rufus}) in New South Wales, Australia. The data $\ygiv$, from \cite[Appendix 8.2; available in Fig.~\ref{fig:data}]{Bayliss:1987}, is a time series of $n=41$ bi-variate observations from double transect counts at irregular time intervals between 1973 and 1984. In \cite{KnapeV12} the authors propose three different models for this data. These models are then compared in a pairwise fashion using the Bayes factor. The comparison has recently been repeated in \cite{ShaoJDT17} using the Hyv\"arinen score. Both \cite{KnapeV12} and \cite{ShaoJDT17} conclude that among the three different models, the preferred model is the following continuous-time stochastic differential equation
\begin{subequations}\label{eq:kangaroo}
	\begin{align}
	x_1 &= \mathcal{LN}(0,5),\\
	\frac{dx_t}{x_t} &= \frac{\sigma^2}{2}dt + \sigma dW_t,\\
	y_{1,t},y_{2,t}\mid x_t &\overset{\text{i.i.d.}}{\sim} \mathcal{NB}\left(x_t,x_t + \tau x_t^2\right),
	\end{align}
\end{subequations}
where $\mathcal{LN}$ is the log-normal distribution, 
$W_t$ is the standard Brownian motion, 
$\mathcal{NB}$ is the negative binomial distribution, and $\param = \{ \sigma, \tau \}$ are unknown parameters. 
The latent variables are $\lat=\{x_t\}_{t\geq 1}$. Note that this model describes a bi-variate time-series $\{ \y_i \}$.

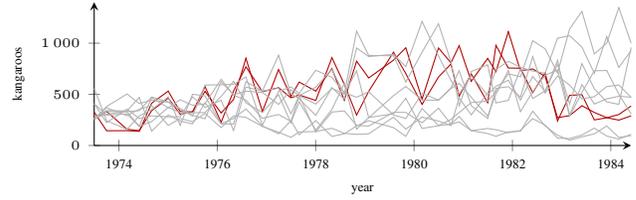
\begin{figure}
%
%
\begin{tikzpicture}

\begin{axis}[%
width=\linewidth,
height=0.4\linewidth,
xmin=1973.497,
xmax=1984.413,
xtick={1974,1976,1978,1980,1982,1984},
xticklabels={1974,1976,1978,1980,1982,1984},
xlabel={year},
ymin=0,
ymax=1400,
ylabel={kangaroos},
axis lines=left,
]
\addplot [color=black!30!red]
  table[row sep=crcr]{%
1973.497	267\\
1973.75	333\\
1974.163	159\\
1974.413	145\\
1974.665	340\\
1975.002	463\\
1975.245	305\\
1975.497	329\\
1975.75	575\\
1976.078	227\\
1976.33	532\\
1976.582	769\\
1976.917	526\\
1977.245	565\\
1977.497	466\\
1977.665	494\\
1978.002	440\\
1978.33	858\\
1978.582	599\\
1978.832	298\\
1979.078	529\\
1979.582	912\\
1979.832	703\\
1980.163	402\\
1980.497	669\\
1980.75	796\\
1980.917	483\\
1981.163	700\\
1981.497	418\\
1981.665	979\\
1981.917	757\\
1982.163	755\\
1982.413	517\\
1982.665	710\\
1982.917	240\\
1983.163	490\\
1983.413	497\\
1983.665	250\\
1983.917	271\\
1984.163	303\\
1984.413	386\\
};

\addplot [color=black!30!red]
  table[row sep=crcr]{%
1973.497	326\\
1973.75	144\\
1974.163	145\\
1974.413	138\\
1974.665	413\\
1975.002	531\\
1975.245	331\\
1975.497	329\\
1975.75	529\\
1976.078	318\\
1976.33	449\\
1976.582	852\\
1976.917	332\\
1977.245	742\\
1977.497	479\\
1977.665	620\\
1978.002	531\\
1978.33	751\\
1978.582	442\\
1978.832	824\\
1979.078	660\\
1979.582	834\\
1979.832	955\\
1980.163	453\\
1980.497	953\\
1980.75	808\\
1980.917	975\\
1981.163	627\\
1981.497	851\\
1981.665	721\\
1981.917	1112\\
1982.163	731\\
1982.413	748\\
1982.665	675\\
1982.917	272\\
1983.163	292\\
1983.413	389\\
1983.665	323\\
1983.917	272\\
1984.163	248\\
1984.413	290\\
};

\addplot [color=white!70!black]
  table[row sep=crcr]{%
1973.497	262\\
1973.75	369\\
1974.163	291\\
1974.413	354\\
1974.665	390\\
1975.002	196\\
1975.245	264\\
1975.497	233\\
1975.75	313\\
1976.078	175\\
1976.33	290\\
1976.582	275\\
1976.917	186\\
1977.245	130\\
1977.497	80\\
1977.665	132\\
1978.002	137\\
1978.33	168\\
1978.582	119\\
1978.832	109\\
1979.078	114\\
1979.582	277\\
1979.832	160\\
1980.163	278\\
1980.497	239\\
1980.75	195\\
1980.917	465\\
1981.163	299\\
1981.497	511\\
1981.665	743\\
1981.917	822\\
1982.163	765\\
1982.413	728\\
1982.665	706\\
1982.917	1049\\
1983.163	1073\\
1983.413	961\\
1983.665	706\\
1983.917	1064\\
1984.163	763\\
1984.413	928\\
};

\addplot [color=white!70!black]
  table[row sep=crcr]{%
1973.497	266\\
1973.75	342\\
1974.163	337\\
1974.413	470\\
1974.665	140\\
1975.002	299\\
1975.245	239\\
1975.497	211\\
1975.75	418\\
1976.078	211\\
1976.33	265\\
1976.582	357\\
1976.917	191\\
1977.245	114\\
1977.497	244\\
1977.665	105\\
1978.002	243\\
1978.33	89\\
1978.582	114\\
1978.832	192\\
1979.078	242\\
1979.582	175\\
1979.832	90\\
1980.163	362\\
1980.497	195\\
1980.75	255\\
1980.917	322\\
1981.163	420\\
1981.497	433\\
1981.665	525\\
1981.917	701\\
1982.163	672\\
1982.413	1024\\
1982.665	945\\
1982.917	708\\
1983.163	1139\\
1983.413	1311\\
1983.665	895\\
1983.917	1000\\
1984.163	1347\\
1984.413	1000\\
};

\addplot [color=white!70!black]
  table[row sep=crcr]{%
1973.497	255\\
1973.75	333\\
1974.163	331\\
1974.413	189\\
1974.665	468\\
1975.002	368\\
1975.245	447\\
1975.497	393\\
1975.75	407\\
1976.078	647\\
1976.33	492\\
1976.582	443\\
1976.917	266\\
1977.245	303\\
1977.497	236\\
1977.665	487\\
1978.002	133\\
1978.33	327\\
1978.582	338\\
1978.832	197\\
1979.078	190\\
1979.582	324\\
1979.832	323\\
1980.163	222\\
1980.497	365\\
1980.75	209\\
1980.917	230\\
1981.163	145\\
1981.497	166\\
1981.665	154\\
1981.917	125\\
1982.163	147\\
1982.413	262\\
1982.665	199\\
1982.917	72\\
1983.163	68\\
1983.413	102\\
1983.665	170\\
1983.917	92\\
1984.163	62\\
1984.413	109\\
};

\addplot [color=white!70!black]
  table[row sep=crcr]{%
1973.497	409\\
1973.75	296\\
1974.163	309\\
1974.413	187\\
1974.665	233\\
1975.002	372\\
1975.245	323\\
1975.497	320\\
1975.75	322\\
1976.078	480\\
1976.33	627\\
1976.582	416\\
1976.917	323\\
1977.245	496\\
1977.497	221\\
1977.665	428\\
1978.002	150\\
1978.33	352\\
1978.582	606\\
1978.832	159\\
1979.078	274\\
1979.582	310\\
1979.832	461\\
1980.163	167\\
1980.497	193\\
1980.75	209\\
1980.917	284\\
1981.163	145\\
1981.497	117\\
1981.665	90\\
1981.917	133\\
1982.163	139\\
1982.413	335\\
1982.665	191\\
1982.917	100\\
1983.163	52\\
1983.413	85\\
1983.665	114\\
1983.917	191\\
1984.163	79\\
1984.413	97\\
};

\addplot [color=white!70!black]
  table[row sep=crcr]{%
1973.497	267\\
1973.75	306\\
1974.163	333\\
1974.413	340\\
1974.665	474\\
1975.002	425\\
1975.245	356\\
1975.497	288\\
1975.75	591\\
1976.078	593\\
1976.33	599\\
1976.582	669\\
1976.917	199\\
1977.245	529\\
1977.497	573\\
1977.665	510\\
1978.002	735\\
1978.33	666\\
1978.582	515\\
1978.832	472\\
1979.078	525\\
1979.582	298\\
1979.832	264\\
1980.163	508\\
1980.497	445\\
1980.75	557\\
1980.917	606\\
1981.163	472\\
1981.497	288\\
1981.665	503\\
1981.917	732\\
1982.163	330\\
1982.413	527\\
1982.665	687\\
1982.917	738\\
1983.163	285\\
1983.413	518\\
1983.665	730\\
1983.917	733\\
1984.163	501\\
1984.413	308\\
};

\addplot [color=white!70!black]
  table[row sep=crcr]{%
1973.497	248\\
1973.75	381\\
1974.163	503\\
1974.413	410\\
1974.665	442\\
1975.002	424\\
1975.245	361\\
1975.497	296\\
1975.75	261\\
1976.078	622\\
1976.33	497\\
1976.582	531\\
1976.917	518\\
1977.245	402\\
1977.497	544\\
1977.665	489\\
1978.002	570\\
1978.33	742\\
1978.582	505\\
1978.832	730\\
1979.078	271\\
1979.582	487\\
1979.832	577\\
1980.163	445\\
1980.497	442\\
1980.75	562\\
1980.917	731\\
1981.163	389\\
1981.497	264\\
1981.665	558\\
1981.917	575\\
1982.163	473\\
1982.413	868\\
1982.665	512\\
1982.917	679\\
1983.163	590\\
1983.413	456\\
1983.665	477\\
1983.917	576\\
1984.163	591\\
1984.413	471\\
};

\addplot [color=white!70!black]
  table[row sep=crcr]{%
1973.497	410\\
1973.75	223\\
1974.163	355\\
1974.413	264\\
1974.665	290\\
1975.002	280\\
1975.245	281\\
1975.497	475\\
1975.75	372\\
1976.078	442\\
1976.33	146\\
1976.582	571\\
1976.917	532\\
1977.245	881\\
1977.497	506\\
1977.665	456\\
1978.002	294\\
1978.33	566\\
1978.582	452\\
1978.832	953\\
1979.078	869\\
1979.582	891\\
1979.832	770\\
1980.163	1212\\
1980.497	875\\
1980.75	637\\
1980.917	505\\
1981.163	680\\
1981.497	494\\
1981.665	579\\
1981.917	638\\
1982.163	444\\
1982.413	676\\
1982.665	527\\
1982.917	832\\
1983.163	296\\
1983.413	595\\
1983.665	405\\
1983.917	456\\
1984.163	481\\
1984.413	472\\
};

\addplot [color=white!70!black]
  table[row sep=crcr]{%
1973.497	557\\
1973.75	243\\
1974.163	157\\
1974.413	289\\
1974.665	351\\
1975.002	474\\
1975.245	198\\
1975.497	501\\
1975.75	485\\
1976.078	211\\
1976.33	210\\
1976.582	624\\
1976.917	532\\
1977.245	560\\
1977.497	311\\
1977.665	486\\
1978.002	405\\
1978.33	558\\
1978.582	503\\
1978.832	1118\\
1979.078	880\\
1979.582	877\\
1979.832	620\\
1980.163	912\\
1980.497	1189\\
1980.75	819\\
1980.917	388\\
1981.163	723\\
1981.497	789\\
1981.665	416\\
1981.917	441\\
1982.163	748\\
1982.413	459\\
1982.665	534\\
1982.917	794\\
1983.163	406\\
1983.413	800\\
1983.665	554\\
1983.917	426\\
1984.163	402\\
1984.413	958\\
};

\end{axis}
\end{tikzpicture}%
	\caption{The kangaroo time series data ($\ygiv$, red), together with a few time series (grey) that are generated from the model in \eqref{eq:kangaroo} (parameters sampled from $w(\param\mid\ygiv)$; cf. Step 7 of Algorithm~\ref{alg:impl}). \dcc indicates no inconsistency between the data and the model ($\apfas = 0.28$), which resonates with the intuition since the generated data behaves `similarly' to the observed data.}
	\label{fig:kangaroo}
\end{figure}

While \cite{KnapeV12} and \cite{ShaoJDT17} favor the model given in \eqref{eq:kangaroo} over other alternatives, we consider the different question whether the model in \eqref{eq:kangaroo} is consistent with the observed data $\ygiv$ or not. We first solve \eqref{eq:kangaroo} analytically to obtain a discrete-time nonlinear state-space model, and use the particle marginal Metropolis-Hastings method \cite{ADH:2010} to sample the unknown parameters. A standard particle filter is used to approximate $p(\y_i\mid \y_1,\dots,\y_{i-1},\param)$. We use $N=1000$ and $M' = M=200$ and obtain $\apfas = 0.28$.  Thus the model in~\eqref{eq:kangaroo} is deemed to be consistent with the observed kangaroo population data (also see Fig.~\ref{fig:kangaroo} for intuitive support of this conclusion).

\section*{Discussion}

We have proposed a data consistency criterion (\dcc) to assess the consistency of a model class $\MC$ with respect to the  observed data $\ygiv$. By comparing the observed (incremental) likelihoods $p(\ygiv_i\mid\ygiv_1, \dots, \ygiv_{i-1},\param)$ to the ones of generated data $\ysim$, \dcc rejects a model class for which $\ygiv$ is atypical for the best models in the class. The criterion follows automatically from the specification of $\MC$ and does not require additional application-specific choices. It yields an (approximate) false alarm probability $\pfas$ of erroneously declaring the best models to be inconsistent. When $\pfas$ falls below some set threshold, there is a sound ground for ruling out the model class~$\MC$. In a sense, the criterion is quantifying the \emph{lack} of inconsistency, see \cite{Robinson:2004} for a discussion about model validation on such grounds.

We have compared \dcc{} to some well-established methods for consistency checks\footnote{Note that model selection methods, such as the Akaike and the Bayesian information criteria, serve a different purpose than \dcc, namely that of relative comparisons between models.}, such as the Kolmogorov-Smirnov and Ljung-Box test. We concluded that \dcc performs, at least, on par with these methods. Moreover, whereas these methods are designed specifically for a certain, rather restricted, model type, \dcc is much more general; indeed \dcc is applicable to a much broader range of classes, ranging from univariate data to high-dimensional time-series models.

By exploiting properties of the Fisher information matrix of $\MC$, it is possible to construct misspecification tests, cf. \cite{White1982_mlemisspecified}. That is, decide whether $p(\y \mid \parambest) = p_0(\y)$ is true or not. Since all practical models are  incomplete in some respect, such tests may not always be relevant.  Instead, what matters in many applications is whether the model is accurate or not, and the proposed \dcc provides a practically useful criterion in this respect.

Using a Bayesian interpretation of \eqref{eq:dcce}, \dcc can be understood as a certain posterior predictive check \cite{R:84,GCS+:14,Box1980_sampling}. Posterior predictive checks are, however, not available for plug-and-play since they require the user to specify a `discrepancy variable', in contrast to the fully automatic \dcc. The \dcc also readily admits a frequentist interpretation in terms of false alarm probability as a sampling property of $p_0(\y)$.

The computational cost of the criterion increases indeed with the dimension of $\param$ and $n$. The operations required are, however, available for many models and well developed in most statistical software packages. \dcc could therefore be used as a routine check in existing statistical modeling methods, in order to better guide the end user to well-grounded scientific conclusions.

\subsection*{Acknowledgments}
This research was financially supported by the Swedish Foundation for Strategic Research (SSF), via the project \emph{ASSEMBLE} (contract number: RIT15-0012), and by the Swedish Research Council, via the project \emph{NewLEADS - New Directions in Learning Dynamical Systems} (contract number: 621-2016-06079).

\bibliography{references}

\begin{figure*}
	\begin{subfigure}{\linewidth}
		\resizebox{\linewidth}{!}{
			\begin{tabular}{@{}cccccccccccccccccccc@{}}
				Magnitude & 1980 & 1981 & 1982 & 1983 & 1984 & 1985 & 1986 & 1987 & 1988 & 1989 & 1990 & 1991 & 1992 & 1993 & 1994 & 1995 & 1996 & 1997 & 1998 \\ \midrule
				$\geq$8 &0&0&0&0&0&2&1&0&0&1&0&0&0&0&2&2&1&0&1 \\
				$\geq$7 &6&10&7&14&14&15&11&13&11&8&18&17&13&12&13&20&15&16&12 \\
				$\geq$6 &96&88&90&139&141&162&140&173&126&139&154&137&179&148&159&201&164&136&121\\
				$\geq$5 &1408&1265&1505&1802&1683&1806&1765&1572&1598&1561&1765&1583&1675&1564&1693&1501&1373&1234&1074
		\end{tabular}}
		
		\vspace{1em}
		
		\resizebox{\linewidth}{!}{
			\begin{tabular}{@{}cccccccccccccccccccc@{}}
				Magnitude & 1999 & 2000 & 2001 & 2002 & 2003 & 2004 & 2005 & 2006 & 2007 & 2008 & 2009 & 2010  & 2011 & 2012 & 2013 & 2014 & 2015 & 2016 & 2017 \\ \midrule
				$\geq$8 &0&1&1&0&1&2&1&2&4&0&1&1&1&2&2&1&1&0&1 \\
				$\geq$7 &18&15&16&13&15&16&11&11&18&12&17&24&20&16&19&12&19&16&7 \\
				$\geq$6 &136&160&137&139&156&157&151&153&196&179&161&175&207&133&142&155&146&146&111\\
				$\geq$5 &1192&1495&1352&1309&1364&1672&1843&1877&2283&1965&2075&2395&2692&1680&1596&1729&1558&1696&1560 
		\end{tabular}}
		\caption{Data for the earthquake count example: The number of earthquakes above a certain magnitude (left column) in the entire world for 1980-2017, retrieved from the U. S. Geological Survey earthquake catalog.}
	\end{subfigure}\\

\vspace{7em}

	\begin{subfigure}{\linewidth}
		\resizebox{\linewidth}{!}{
			\begin{tabular}{@{}ccccccccccccccc@{}}
				Date&1973 Jul&1973 Oct&1974 Mar&1974 Jun&1974 Sep&1975 Jan&1975 Apr&1975 Jul&1975 Oct&1976 Feb&1976 May&1976 Aug&1976 Dec&1977 Apr \\ \midrule
				Counts &267&333&159&145&340&463&305&329&575&227&532&769&526&565 \\
				&326&144&145&138&413&531&331&329&529&318&449&852&332&742
		\end{tabular}}
		
		\vspace{1em}
		
		\resizebox{\linewidth}{!}{
			\begin{tabular}{@{}ccccccccccccccc@{}}
				Date & 1977 Jul&1977 Sep&1978 Jan&1978 May&1978 Aug&1978 Nov&1979 Feb&1979 Aug&1979 Nov&1980 Mar&1980 Jul&1980 Oct&1980 Dec&1981 Mar \\ \midrule
				Counts &466&494&440&858&599&298&529&912&703&402&669&796&483&700 \\
				&479&620&531&751&442&824&660&834&955&453&953&808&975&627
		\end{tabular}}
		
		\vspace{1em}
		
		\resizebox{\linewidth}{!}{
			\begin{tabular}{@{}cccccccccccccc@{}}
				Date & 1981 Jul&1981 Sep&1981 Dec&1982 Mar&1982 Jun&1982 Sep&1982 Dec&1983 Mar&1983 Jun&1983 Sep&1983 Dec&1984 Mar&1984 Jun \\ \midrule
				Counts &418&979&757&755&517&710&240&490&497&250&271&303&386 \\
				&851&721&1112&731&748&675&272&292&389&323&272&248&290
		\end{tabular}}
		\caption{Data for the kangaroo population example, adopted from \citet[Appendix 8.2]{Bayliss:1987}}
	\end{subfigure}
\caption{Complete data sets for the earthquake and kangaroo counting examples.}
\label{fig:data}
\end{figure*}

\end{document}